\documentclass[twocolumn]{aastex63}

\definecolor{purple}{rgb}{0.5,0,0.87}
\definecolor{teal}{rgb}{0,0.67,0.67}
\definecolor{lightblue}{rgb}{0.1,0.5,0.89}
\definecolor{mblue}{rgb}{0.35, 0.32, 1} 
\definecolor{pinky}{rgb}{0.8,0.04,0.8} 

\hypersetup{linkcolor=teal, citecolor=pinky, filecolor=cyan, urlcolor=pinky}

\newcommand{\dft}{NGC~1052--DF2}
\newcommand{\dff}{NGC~1052--DF4}
\newcommand{\kks}{KKS2000[04]}
\newcommand{\hipercam}{HiPERCAM}
\newcommand{\sextractor}{\texttt{SExtractor}}
\newcommand{\scamp}{\texttt{SCAMP}}
\newcommand{\swarp}{\texttt{SWarp}}
\newcommand{\ellipse}{\texttt{ellipse}}
\defcitealias{vD2018}{vD18}
\defcitealias{Montes2020}{M20}

\received{January 1, 2018}
\revised{January 7, 2018}
\accepted{}

\submitjournal{ApJ}

\shorttitle{A disk and no tidal features in DF2}
\shortauthors{Montes et al.}

\begin{document}

\title{A disk and no signatures of tidal distortion in the galaxy ``lacking" dark matter \dft}

\correspondingauthor{Mireia Montes}
\email{mireia.montes.quiles@gmail.com}

\author[0000-0001-7847-0393]{Mireia Montes}
\altaffiliation{STScI Prize Fellow}
\affiliation{Space Telescope Science Institute, 3700 San Martin Drive, Baltimore, MD 21218, USA}

\author[0000-0001-8647-2874]{Ignacio Trujillo}
\affiliation{Instituto de Astrof\'{\i}sica de Canarias, c/ V\'{\i}a L\'actea s/n, E38205 - La Laguna, Tenerife, Spain}
\affiliation{ Departamento de Astrof\'isica, Universidad de La Laguna, E-38205 - La Laguna, Tenerife, Spain}   

\author[0000-0002-6220-7133]{Ra\'ul Infante-Sainz}
\affiliation{Instituto de Astrof\'{\i}sica de Canarias, c/ V\'{\i}a L\'actea s/n, E38205 - La Laguna, Tenerife, Spain}
\affiliation{ Departamento de Astrof\'isica, Universidad de La Laguna, E-38205 - La Laguna, Tenerife, Spain}

\author[0000-0001-5292-6380]{Matteo Monelli}
\affiliation{Instituto de Astrof\'{\i}sica de Canarias, c/ V\'{\i}a L\'actea s/n, E38205 - La Laguna, Tenerife, Spain}
\affiliation{ Departamento de Astrof\'isica, Universidad de La Laguna, E-38205 - La Laguna, Tenerife, Spain}

\author[0000-0003-3249-4431]{Alejandro S. Borlaff}
\affiliation{NASA Ames Research Center, Moffett Field, CA 94035, USA}

\begin{abstract}
Using ultra-deep imaging ($\mu_g = 30.4$ mag/arcsec$^2$; 3$\sigma$, 10$\arcsec\times$10\arcsec), we probed the surroundings of the first galaxy ``lacking" dark matter \kks{} (\dft). Signs of tidal stripping in this galaxy would explain its claimed low content of dark matter. However, we find no evidence of tidal tails. In fact, the galaxy remains undisturbed down to a radial distance of $80$ arcsec. This radial distance triples previous spatial explorations of the stellar distribution of this galaxy.  In addition, the distribution of its globular clusters (GCs) is not extended in relation to the bulk of the galaxy (the radius containing half of the GCs is $21$ arcsec). We also found that the surface brightness radial profiles of this galaxy in the $g$ and $r$ bands decline exponentially from 35 to 80 arcsec. That, together with a constant ellipticity and position angle in the outer parts of the galaxy strongly suggests the presence of a low-inclination disk. This is consistent with the evidence of rotation found for this object. This finding implies that the dynamical mass of this galaxy is a factor of 2 higher than previously reported, bringing the dark matter content of this galaxy in line with galaxies of similar stellar mass. 

\end{abstract}

\keywords{editorials, notices --- 
miscellaneous --- catalogs --- surveys}


\section{Introduction}\label{sec:intro}

Dark matter (DM) is a key constituent in the current favored models of galaxy formation. For this reason, the claimed existence of two old and long-lived galaxies ``lacking'' DM \citep{vD_df2, vD2019} in the field of view (FOV) of the massive elliptical galaxy NGC~1052 sparked a lot of controversy \citep[][ to name a few]{Martin2018, Trujillo2019, Laporte2019, Monelli2019, Nusser2019}. These galaxies have been argued to have a stellar vs. dark matter mass ratio of $\sim1$, while for similar galaxies the observed ratios are $\lesssim 0.1$ \citep[e.g.,][]{Mancera2019, Mueller2020}. Their existence could imply a challenge to the current galaxy formation paradigm, as without DM the gas would not have collapsed and formed stars.

A growing number of theoretical studies suggest an explanation for the existence of these galaxies within the current cosmological paradigm: tidal stripping of the DM of these galaxies \citep[][]{Ogiya2018, Nusser2020, Yang2020, Jackson2020, Maccio2020, Doppel2021}.  In a recent paper, \citet[][hereafter, \citetalias{Montes2020}]{Montes2020} found that the second galaxy ``lacking" DM, \dff, is showing tidal tails pointing to an interaction with its nearby neighbor, NGC 1035. As stars are more centrally concentrated than the DM, this means that tidal stripping has removed a significant fraction of the DM before starting to affect the stars of the galaxy.

The current cosmological paradigm, Lambda Cold Dark Matter ($\Lambda$CDM), predicts significant numbers of faint stellar streams in the outskirts of the majority of nearby massive galaxies \citep[e.g.,][]{Bullock2005, Cooper2010}. All this structure starts to show up at surface brightness fainter than $30$ mag/arcsec$^2$. However, the detection and characterization of faint substructures in stellar halos is still a work in progress as their intrinsic low surface brightness prevents a complete census of these structures.

In this paper, we take advantage of ultra-deep observations and recent advances in low surface brightness data processing to explore the tidal stripping scenario in \kks{}, better known as \dft{} (for convenience, this is the name we will use throughout this paper). The presence of tidal tails surrounding \dft{} will serve as the smoking gun of ongoing tidal disruption and therefore, explain the ``lack” of DM observed in the first galaxy of this kind. We follow the same steps than in \citetalias{Montes2020} to search for signs of interaction in this galaxy. Sec. 2 presents the data used in this work. Then, we explore the distribution of GCs in Sec. \ref{sec:gcs} as tidal stripping will place them along the orbit of the galaxy. Secondly, we explore the stellar light of the galaxy, deriving its surface brightness profiles in Sec. \ref{sec:profiles} and its shape in Sec. \ref{sec:shape}. Finally, we discuss the implications of our findings for the claimed rotation of the galaxy (Sec. \ref{sec:rotation}) and its association with its closest (in projection) neighbors (Sec. \ref{sec:disc_tidal}). All magnitudes in this work are given in the AB magnitude system.


\section{Data}

The data used in this paper come from three different facilities: the \emph{Hubble Space Telescope} (\emph{HST}) and two ground based telescopes: the 10.4m Gran Telescopio Canarias (GTC) and the 2.5m Isaac Newton Telescope (INT). We describe the details of each observation on what follows.

\subsection{Hubble Space Telescope imaging}

\dft{} was observed with \emph{HST} ACS Wide-Field Channel (WFC) as part of the program GO-14644 (PI: van Dokkum) available through the MAST archive\footnote{\url{https://mast.stsci.edu/portal/Mashup/Clients/Mast/Portal.html}}. The data consists of one orbit in the $F606W$ and one in the $F814W$ band. The total exposure time is $545$\,s in $F606W$ and $580$\,s in $F814W$.

The data reduction of the \emph{HST} imaging of \dft{} is described in detail in \citet{Trujillo2019} and summarized below. 
The charge-transfer efficiency (CTE) corrected data were downloaded from MAST and \texttt{flc} files were used to build the final drizzled mosaics with \texttt{Astrodrizzle} \citep{Gonzaga2012}. To estimate the sky, we masked thoroughly the images using a combination of \texttt{NoiseChisel} \citep{Akhlaghi2015} and manual masking of the galaxy. The centroid of the sky distribution derived using the masked images was calculated using bootstrapping\footnote{Using the code \texttt{Bootmedian}, available at \url{https://github.com/Borlaff/bootmedian}}. Finally, to mitigate the gradients present in the images, \texttt{NoiseChisel} was used to generate a final sky background model of the final co-added images that was subtracted from the images. 

Following \citetalias{Montes2020}, we use these high-resolution \emph{HST} observations together with photometry from the GTC \hipercam{} deep data to characterize and potentially find new GCs around \dft.

\subsection{Deep ground-based imaging}

One of the biggest challenges in low surface brightness imaging is to process ultra-deep observations in a manner that preserves the faint light of the object of interest while correcting for all systematic effects introduced by the instrument/telescope. In this section, we describe the data reduction performed for our ground-based ultra-deep imaging. Both the data acquisition and data processing are aimed at preserving the low surface brightness light around \dft. Most of the reduction steps are common for both datasets and described in the following. Specific steps for each of the telescopes are discussed later. 

A key step for obtaining ultra-deep images is the dithering pattern. \dft{} was observed by the GTC and INT following the dithering strategy outlined in \citet{Trujillo2016} aimed at reducing scattered light from the telescope structure as much as possible. In short, this consists in a dithering pattern with large steps (typically the size of the source under investigation) and whenever possible rotation of the camera. This ensures we achieve the flat background required for our goals.

Bias frames were taken the same night of the observations as part of the different observing programs. The master bias frames were created as the $3\sigma$-clipped mean of the individual frames and subtracted from the original images. The flat-field frames are derived using the same science exposures obtained for this work\footnote{Dome or twilight flats can introduce gradients in the data due to inhomogeneities in illumination. Using the same science images to derive the flat-field frames ensures that we are not introducing gradients \citep[e.g.,][]{Trujillo2016}.}. The bias corrected CCD frames present steep gradients making the proper masking of the sources difficult in those science frames, a crucial step to achieve an accurate flat-field correction. For this reason, the flat-field frames were built following a two-step process. First, we derive a preliminary flat by stacking each bias-corrected frame. This preliminary flat is used to correct each CCD frame in order to flatten and, consequently, properly mask the science exposures. Second, we combine these masked, and normalized, images to create the final master flat-field frames.

We, then, performed the astrometric calibration of the different frames. We used the \texttt{astrometry.net} software \citep{Lang2010} to produce an approximate astrometric solution, later refining it with \scamp{} \citep{Bertin2006}. 

Once all frames were corrected of all systematic effects and in a common astrometric solution, they were resampled into a common grid using \swarp{} \citep{Bertin2010} and stacked using a $3\sigma$-clipped mean into a final image, one per filter.

\subsubsection{GTC \hipercam{} images}

\hipercam{} \citep{Dhillon2018} is a quintuple-beam, high-speed astronomical imager able to obtain images of celestial objects in five different filters ($u$, $g$, $r$, $i$, $z$) simultaneously. The image area of each of the five CCDs is $2048\times1024$ pixels ($2.7\arcmin \times1.4\arcmin$; 1 pixel$=0\farcs08$) divided into four channels of $1024\times512$ pixels. \dft{} was observed on the 2019-01-11. The whole reduction process was done within a controlled and enclosed software environment as described in \citet{Akhlaghi2020}.

The data reduction of the \dft{} images follows the steps detailed in \citetalias{Montes2020} for \dff. After the standard calibration per CCD channel (bias and flat-field), each set of four channels were assembled into a single image. The different exposures that went into the final images were visually inspected and those with low quality or strong gradients were rejected. The photometric calibration of these images was done using SDSS DR12 \citep{Alam2015}. The final exposure time is 0.6h for each band, corresponding to point-like source depths ($5\sigma$ in $2\arcsec$ diameter apertures) of: $25.32 \pm 0.05$, $24.69\pm0.05$, $24.63\pm0.07$, $24.18\pm0.06$, $23.71\pm0.06$, in the $u$, $g$, $r$, $i$, $z$ bands, respectively. The sky subtraction was done by subtracting a constant value computed  from the masked image.

Unfortunately, the FOV of \hipercam{} is comparable to the size of the galaxy making it impossible to reliably measure the sky. This results in over-subtraction in the outer parts of the galaxy. Therefore, we will use these images to study only the photometry of the GC system around \dft.

\subsubsection{INT images}

The Isaac Newton Telescope (INT) is a 2.5-m telescope situated in the Roque de los Muchachos Observatory in the island of La Palma. This telescope is equipped with a wide-field camera (WFC), mounted at the prime focus, providing imaging over a 33-arcmin FOV with a wide variety of broad- and narrow-band filters. The WFC consists of 4 thinned EEV 2k$\times$4k CCDs. The pixel scale of the camera is $0\farcs33$. The data were taken between 2020-10-12 and 2020-10-16 in two different filters $g$ and $r$ for a total exposure time of  4.4h and 3.2h, respectively.

In this case, after following the data reduction outlined above, the photometric calibration of the images was done using the SDSS DR15 \citep{Aguado2019}. Thanks to the high quality of the flat-field correction, the sky subtraction was done by only using an average constant applied to the entire image. No gradients were present in the final co-adds of each band. 

The nominal depth of the final images measured in $10\arcsec\times10\arcsec$ boxes is $30.4$ and $29.5$ mag/arcsec$^2$ in the $g$ and $r$ bands ($3\sigma$ above the background), respectively, measured using the method in appendix A in \citet{Roman2020}. For comparison, the limiting surface brightness (3$\sigma$; 12\arcsec$\times$12\arcsec) of the Dragonfly images in the field of NGC1052 are between $\mu_{g/r} =$ 27.4 and 28.0 mag/arcsec$^2$ \citep[][]{Merritt2016} and $\mu_r =$ 28.5 mag/arcsec$^2$ in \citet[][]{Muller2019} ($3\sigma$; $10\arcsec\times10\arcsec$).

\subsection{Scattered light removal from nearby stars}\label{sec:rem_star}
The careful modeling and removal of stars in deep images is now a common technique in low surface brightness science \citep[e.g, ][]{Slater2009, Trujillo2016, Infante2020, Montes2021}. In order to reliably explore low surface brightness features around \dft{}, it is key to remove the light scattered by the brightest objects in the image. In fact, there is the bright star HD16873 (R.A. = $2^{h}42^{m}9.3^{s}$, Dec = $-8^{d}22\arcmin32\farcs218$, V= 8.34 mag), $\sim350$ arcsec to the east of \dft, creating a source of extra light, contamination, that can bias our results. The removal of this scattered light contamination is particularly imperative for the images where we are exploring the distribution of light around \dft{}: the INT $g$ and $r$ images. 

To correct for this scattered light, we first need to model the 2D profile of the star. We use the software \ellipse{} in IRAF. \ellipse{} fits elliptical isophotes to the 2-D images of sources using the method described in \citet{Jedrzejewski1987}. 

Prior to modelling the star, we need to carefully mask foreground and background sources as their presence can affect the fitting. For this, we used a two-step \sextractor{} setting; a ``hot+cold” mode \citep[][]{Rix2004, Montes2021}. The ``cold" mode masks extended sources while the ``hot” mode is optimized to detect the more compact and faint sources that are embedded within the galaxy light.
We run \sextractor{} in a deep combined INT $g+r$ image. The ``cold” mask was further expanded 8 pixels while the “hot” was expanded 2 pixels, leaving the bright star we want to model unmasked. Because the PSF in our INT images is highly asymmetric, especially noticeably in bright stars, we only modeled the section of the star that is close to \dft{}, masking the rest manually. This way we ensure that the output model is suitable for our purposes. In addition, we manually masked \dft{} to cover any diffuse light that has remained unmasked from the above procedure that could contaminate the flux of the star at large radius.

Once \ellipse{} was run in the masked image, we created the model of the star with \texttt{bmodel} and subtracted it from the images. In addition, we also modelled and subtracted two fainter stars $\sim150$ arcsec to the east and west of \dft{} (R.A. = $2^{h} 41^{m}36.88^s$ Dec = $-8^d24\arcmin56\farcs80$, and R.A. = $2^h41^m54.47^s$ Dec =  $-8^d24\arcmin20\farcs80$) .


\section{The spatial distribution of the globular cluster system of \dft}\label{sec:gcs}

Due to the compact sizes (half-light radii of a few parsecs) and masses of globular clusters \citep[GCs, $10^4 - 10^6$ M$_\odot$;][]{Harris1996}, they appear as good tracers of the properties of their host galaxy as they are easily observable in distant system even at large radial distances \citep[see][]{Brodie2006}. 

The spatial distribution of the GC system of a galaxy reveals clues about its present state. In an accretion event, the stripped material is placed along the orbit of the parent satellite galaxy, forming tidal streams \citep[e.g., ][]{Toomre1972, Johnston1996}. In the same way, the GCs will also be placed along the orbit and exploring their observed spatial distribution can indicate if this process is happening and the direction of the orbit.

One of the goals of this work is to explore whether the GCs align through a particular direction, suggesting that \dft{} is undergoing tidal stripping or, on the contrary, the GCs are more spherically distributed around \dft{}, which is expected if the galaxy is not experiencing any interaction.

\subsection{GC selection}

 \begin{figure*}
 \centering
   \includegraphics[width = 0.9\textwidth]{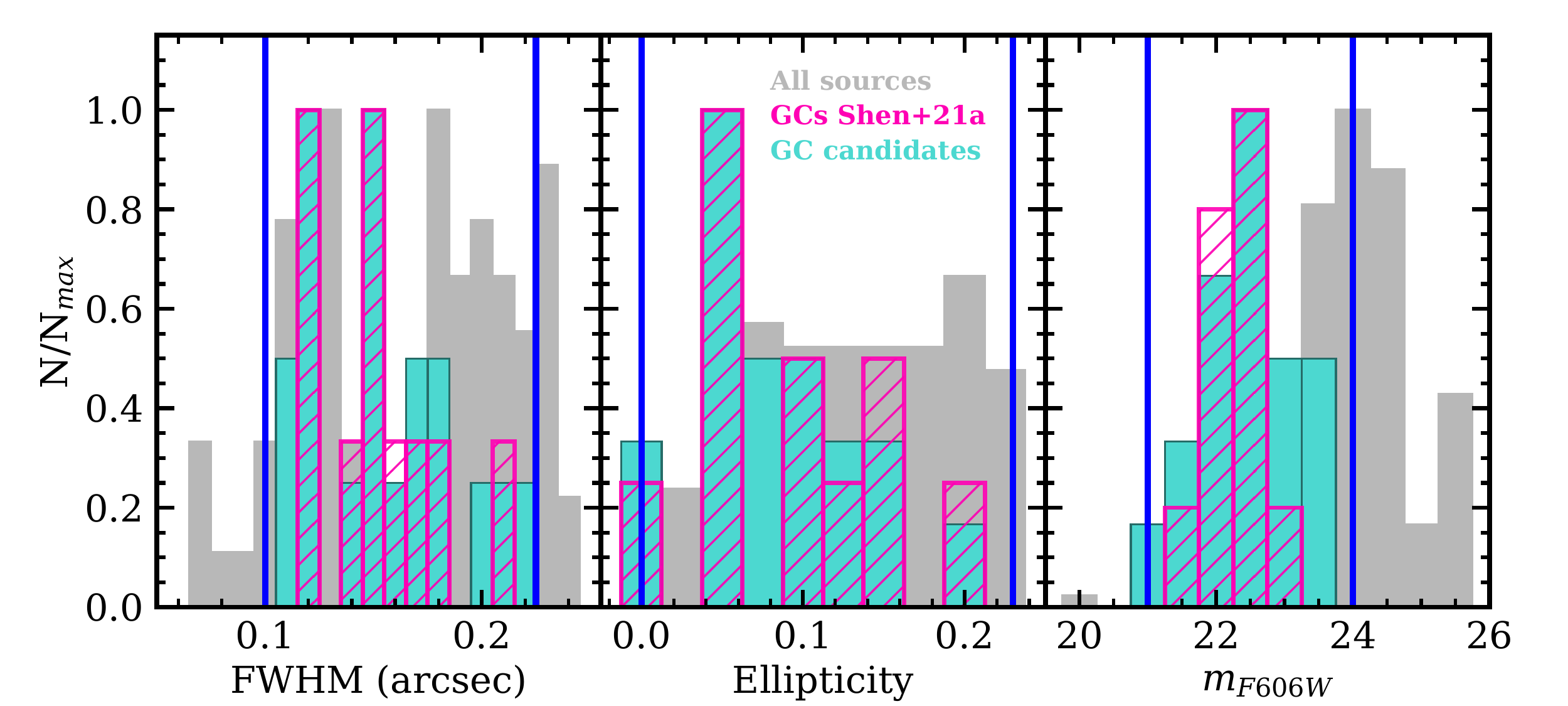}
   \caption{Histograms of FWHM (left), ellipticity (middle) and $F606W$ magnitude m$_{F606W}$ (right) for all the sources simultaneously detected sources in the HST and \hipercam{} images (grey). The magenta histograms show the distribution of the spectroscopically confirmed GCs in \citet{Shen2020}. The blue vertical lines mark the range in each parameter of the initial selection. The mint green histograms show the sources that simultaneously satisfy the conditions of FWHM, ellipticity and magnitude enclosed by the vertical blue lines.\label{fig:gc_selection}}
    \end{figure*}

In this section, we combine the high resolution of \emph{HST} with the multicolor information from \hipercam{} to identify new GCs for \dft. Recent works have shown the increased effectiveness of identifying GCs using more information from their spectral energy distributions \citep[][]{Montes2014, Munoz2014}. 

\citet{vD_df2} and \citet{Shen2020} confirmed spectroscopically 12 GCs in \dft. 
Here, we try to identify more GCs following the steps in \citetalias{Montes2020}. First, we run \sextractor{} \citep{Bertin1996} on the \emph{HST} images in dual-mode using F814W as the detection image. In the same way, \sextractor{} was also run in the \hipercam{} images using $r$ as our detection image. Both catalogues were matched based on the position of the sources in the sky. 

We pre-selected GC candidates thanks to the high-resolution of the \emph{HST} data\footnote{We have rerun the analysis with the images of \dft{} from GO-15851 (PI: van Dokkum) and the results obtained do not change with this deeper dataset.}. This selection is based upon the properties of the confirmed GCs in \citet{Shen2020}. In addition to the morphology selection with \emph{HST} similar to that in \citetalias{Montes2020}, we also used the measured $F606W - F814W$ color to eliminate sources that are clearly too red to be a GC candidate.

The pre-selection criteria is as follows.

\begin{itemize}
    \item $0.1\arcsec<$ FWHM $<0.23\arcsec$
    \item $0<$ ellipticity $<0.23$
    \item $21<$ m$_{F606W}$ $<24$ mag 
    \item m$_{F606W}$ - m$_{F814W}$ $<0.8$  mag
\end{itemize}

 \begin{figure}
 \centering
   \includegraphics[width = 0.45\textwidth]{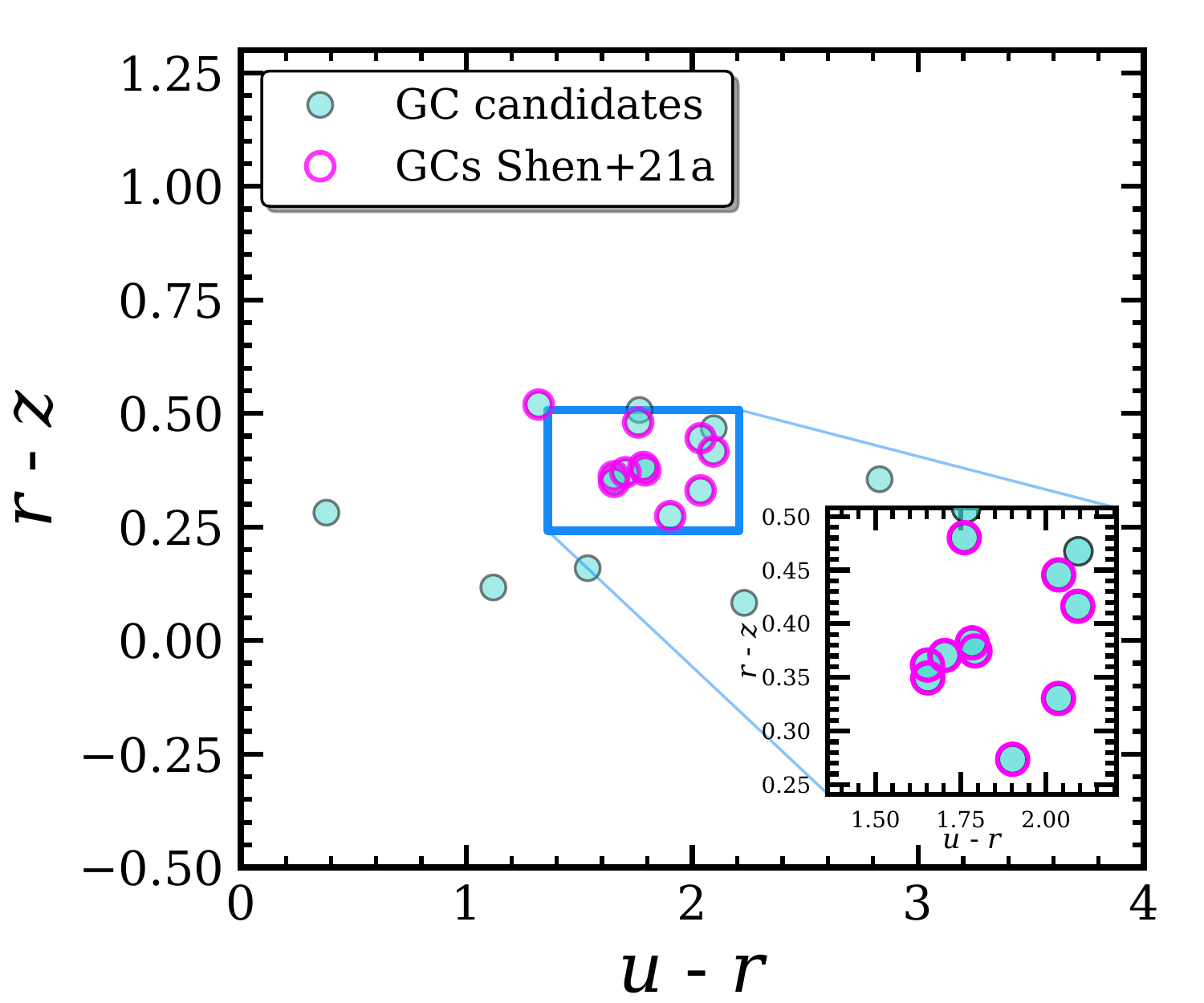}
   \caption{The ($u$-$r$)-($r$-$z$) color-color diagram of the initial sample of candidate GCs (mint green). The spectroscopically confirmed GCs of \citet{Shen2020} are highlighted in magenta. The blue box indicates the color-color region we have selected to create our final sample of GC candidates. The inset shows a zoom-in into the selection box for ease of viewing.
   \label{fig:gc_colors}}
    \end{figure}
    

The aim of this selection is to find GCs that are similar to the confirmed GCs while allowing for sources that might have been too faint for spectroscopic follow-up.
Fig. \ref{fig:gc_selection} shows the histograms of FWHM, ellipticity and m$_{F606W}$ of all the sources detected with \sextractor{} (filled in grey), the confirmed GCs in \citet{Shen2020} (dashed in magenta) and the GC candidates based in our preliminary selection (filled in mint green). Nineteen objects fall within this pre-selection, including eleven objects in \citet{Shen2020} (see below).

The next step is using the colors of the GCs from \hipercam{} to narrow down the selection. Fig. \ref{fig:gc_colors} shows the ($u$-$r$) vs. ($r$-$z$) color-color diagram for the pre-selected GC candidates (mint green) from our preliminary selection. We have marked 11 GCs in \citet{Shen2020} as the magenta circles\footnote{One of the GCs identified in \citet{Shen2020} (GC-39) is outside the FOV of \hipercam, and has been added manually after this step.}. As seen in Fig. \ref{fig:gc_colors}, the confirmed GCs fill a very narrow region in the color-color space and, therefore, we further narrowed down the initial selection to objects within $1.4<u-r<2.2$ mag and $0.24<r-z<0.51$ mag (blue box). These ranges in color are based on the $\pm2\sigma$ around the median colors of the confirmed GCs (magenta circles), with $\sigma$ being their dispersion in color. One of the spectroscopically confirmed GCs (C-14) fall outside this selection box as it appears to be blended with another source and therefore its photometry might be contaminated, but was not discarded from the final selection. 

We identified fourteen GCs, two in addition to the ones confirmed spectroscopically. The coordinates and magnitudes of all the objects are listed in Table \ref{table:gcs}. The magnitudes are corrected by the extinction of our Galaxy \citep{Schlafly2011}: A$_u = 0.10$, A$_g = 0.08$, A$_r = 0.06$, A$_z = 0.03$, A$_{F606W} = 0.06$ and A$_{F814W} = 0.04$. Two of the GCS (C-6 and C-10) have ground-based magnitudes that are brighter than their \emph{HST} magnitudes suggesting that they are affected by blending.

 \begin{deluxetable*}{llcccccccc}[t]
 \tabcolsep=0.1cm
\tablecaption{\label{table:gcs}
	Confirmed and candidate GCs of \dft.}
\tablehead{ ID & Other ID$^a$ & RA & DEC & $m_u$ & $m_g$ & $m_r$ &  $m_z$ & $m_{\rm F606W}$ & $m_{\rm F814W}$ \\
&  & &  & $\mathrm{mag}$ & $\mathrm{mag}$ & $\mathrm{mag}$ & $\mathrm{mag}$ & $\mathrm{mag}$ & $\mathrm{mag}$ }
\startdata
C-1$^*$& GC-59 &2$^h$41$^m$48.09$^s$ & $-8\arcdeg24\arcmin56.82\arcsec$ & $24.23 \pm 0.28$ & $22.86 \pm 0.04$ & $22.58 \pm 0.03$ & $22.22 \pm 0.07$ & $22.82 \pm 0.01$ & $22.38 \pm 0.01$\\
C-2$^*$ & GC-71 &2$^h$41$^m$45.14$^s$  & $-8\arcdeg24\arcmin22.31\arcsec$ & $24.51 \pm 0.20$ & $22.79 \pm 0.03$ & $22.47 \pm 0.03$ & $22.14 \pm 0.06$ & $22.66 \pm 0.01$ & $22.26 \pm 0.01$ \\
C-3$^*$ & GC-73 &2$^h$41$^m$48.22$^s$  & $-8\arcdeg24\arcmin17.49\arcsec$ & $22.96 \pm 0.06$ & $21.55 \pm 0.01$ & $21.31 \pm 0.01$ & $20.96 \pm 0.02$ & $21.49 \pm 0.01$ & $21.15 \pm 0.01$ \\
C-4   &  & 2$^h$41$^m$47.61$^s$  & $-8\arcdeg24\arcmin18.21\arcsec$ & $25.39 \pm 0.61$ & $23.87 \pm 0.08$ & $23.63 \pm 0.07$ & $23.12 \pm 0.15$ & $23.73 \pm 0.03$ & $23.33 \pm 0.03$  \\
C-5$^*$ & GC-77 & 2$^h$41$^m$46.55$^s$  & $-8\arcdeg24\arcmin13.36\arcsec$ & $23.54 \pm 0.09$ & $21.96 \pm 0.02$ & $21.78 \pm 0.01$ & $21.30 \pm 0.03$ & $21.98 \pm 0.01$ & $21.61 \pm 0.01$ \\
C-6$^*$ & GC-85 & 2$^h$41$^m$47.75$^s$  & $-8\arcdeg24\arcmin05.30\arcsec$ & $20.89 \pm 0.13$ & $19.42 \pm 0.02$ & $19.10 \pm 0.02$ & $18.73 \pm 0.04$ & $22.39 \pm 0.01$ & $21.97 \pm 0.01$  \\
C-7 &    & 2$^h$41$^m$45.91$^s$  & $-8\arcdeg23\arcmin56.43\arcsec$ & $25.77 \pm 0.73$ & $24.09 \pm 0.09$ & $23.67 \pm 0.08$ & $23.20 \pm 0.18$ & $23.82 \pm 0.03$ & $23.39 \pm 0.03$  \\
C-8$^*$& GC-91 & 2$^h$41$^m$42.17$^s$  & $-8\arcdeg23\arcmin53.27\arcsec$ & $24.20 \pm 0.25$ & $22.43 \pm 0.02$ & $22.16 \pm 0.02$ & $21.72 \pm 0.05$ & $22.44 \pm 0.01$ & $22.04 \pm 0.01$  \\
C-9$^*$& GC-92 &2$^h$41$^m$46.89$^s$  & $-8\arcdeg23\arcmin50.69\arcsec$ & $23.32 \pm 0.12$ & $21.96 \pm 0.02$ & $21.61 \pm 0.02$ & $21.24 \pm 0.04$ & $22.35 \pm 0.01$ & $21.83 \pm 0.01$ \\
C-10$^*$& GC-93 &2$^h$41$^m$46.72$^s$ & $-8\arcdeg23\arcmin50.75\arcsec$ & $21.87 \pm 0.26$ & $20.08 \pm 0.04$ & $19.78 \pm 0.03$ & $19.36 \pm 0.07$ & $22.91 \pm 0.02$ & $22.53 \pm 0.02$ \\
C-11$^*$ & GC-98 & 2$^h$41$^m$47.34$^s$ & $-8\arcdeg23\arcmin34.54\arcsec$ & $24.44 \pm 0.24$ & $22.90 \pm 0.03$ & $22.65 \pm 0.03$ & $22.27 \pm 0.07$ & $22.85 \pm 0.01$ & $22.44 \pm 0.01$  \\
C-12$^*$ & GC-101 & 2$^h$41$^m$45.20$^s$ & $-8\arcdeg23\arcmin27.73\arcsec$ & $24.56 \pm 0.28$ & $22.95 \pm 0.04$ & $22.66 \pm 0.04$ & $22.39 \pm 0.09$ & $22.98 \pm 0.02$ & $22.54 \pm 0.01$  \\
C-13$^*$ & GC-39 & 2$^h$41$^m$45.07$^s$ & $-8\arcdeg25\arcmin24.87\arcsec$ & ... & ...  & ...  & ...  & $22.31 \pm 0.01$ & $21.93 \pm 0.01$  \\
C-14$^*$& GC-80 & 2$^h$41$^m$46.23$^s$ & $-8\arcdeg24\arcmin07.32\arcsec$ & $23.54 \pm 0.12$ & $22.35 \pm 0.03$ & $22.22 \pm 0.02$ & $21.70 \pm 0.05$ & $22.56 \pm 0.01$ & $ 22.17 \pm 0.01$ \\
\hline
\multicolumn{10}{c}{$^*$GC spectroscopically confirmed in \citet{Shen2020}. } \\
\multicolumn{10}{c}{$^a$ ID in \citet{Shen2020}.} 
\enddata
\end{deluxetable*}

\subsection{Spatial distribution of the GCs}

In this section, we address the spatial distribution of the GCs in \dft{} in order to obtain clues about the interaction state of the galaxy. As a satellite galaxy is interacting with a massive galaxy, the stripped material (both stars and GCs) of the satellite will be deposited along its orbit. This is the case for \dff{} where its GC system align in a particular axis along the galaxy (see Fig. 4 in \citetalias{Montes2020}) and the spatial distribution of GCs is more extended than that of the stellar body of the galaxy.

\begin{figure}
 \centering
   \includegraphics[width = 0.45\textwidth]{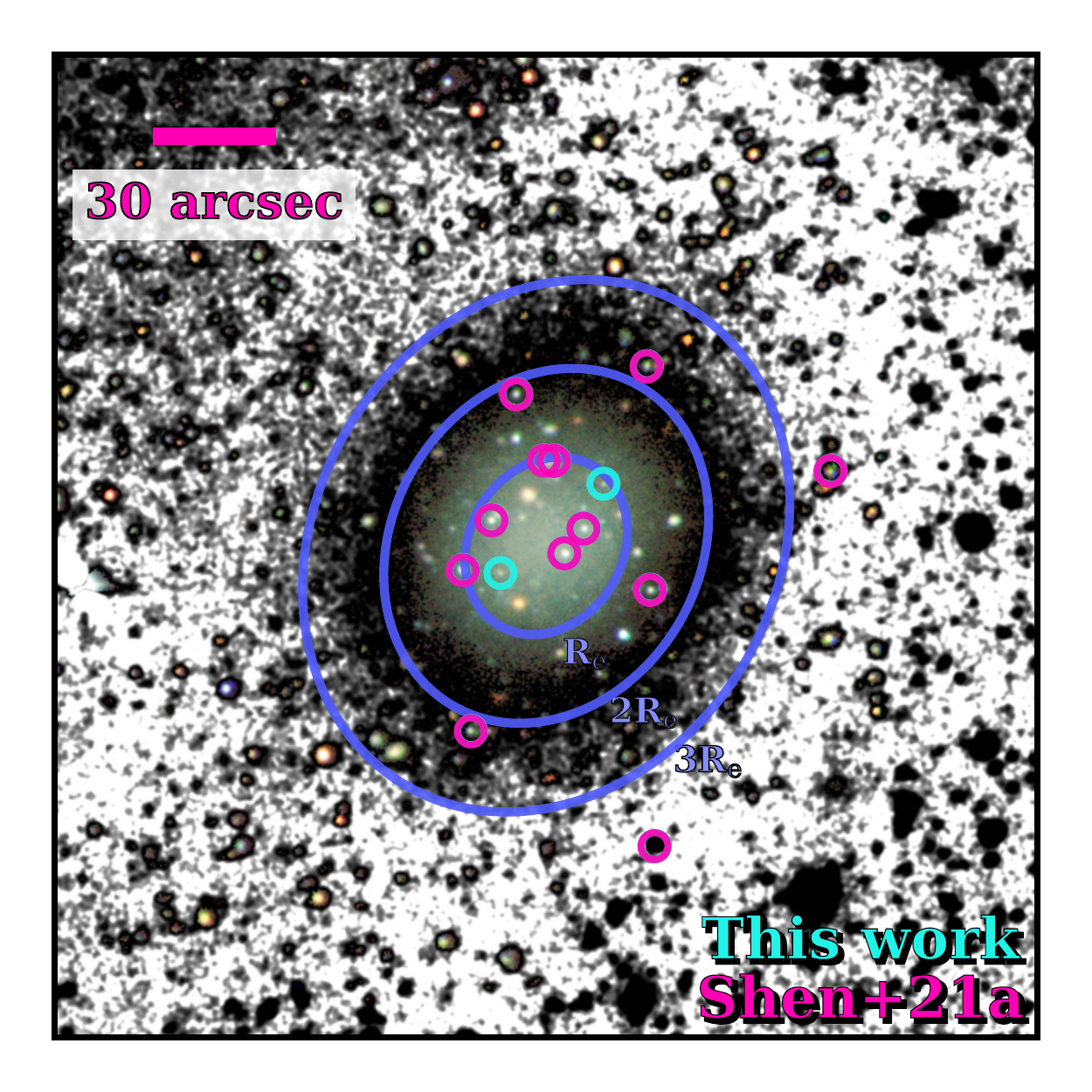}
   \caption{Color composite of a postage stamp of a region $240\arcsec\times240\arcsec$ around \dft{}. The black and white background is the $g$-band INT image. The mint green circles highlight the candidate GCs found in this work while the magenta circles are the GCs spectroscopically confirmed in \citet{Shen2020}. The blue ellipses indicate $1R_e$, $2R_e$ and $3R_e$, respectively, of the stellar light of the galaxy ($R_e = 22.6$ arcsec, as reported in \citealt{vD_df2}).
   \label{fig:gc_distribution}}
    \end{figure}

Fig. \ref{fig:gc_distribution} shows an RGB image created using the \hipercam{} $g$, $r$ and $z$ and a black and white INT $g$ image for the background. The highlighted sources are the 12 confirmed GCs from \citet[][magenta]{Shen2020} and the two new candidates obtained in the previous section (mint green). From Figure \ref{fig:gc_distribution}, it is evident that the distribution of GCs of \dft{} is remarkably different from that of \dff; the spatial distribution of the GCs in \dft{} is more concentrated. The radius containing half of the GCs (R$_{e, GC}$) is $21.0\pm0.1$ arcsec. This is similar to the value of the effective radius published for the stellar light distribution of this galaxy using HST data \citep[$R_e = 22.6$ arcsec;][]{vD_df2}. In contrast, the radius containing half of the GCs in \dff{} was 9 times larger than the $R_e$ of the stellar light (\citetalias{Montes2020}).
Furthermore, all but two of the GCs are contained in the stellar body of the galaxy ($<3R_e$). This concentrated distribution of GCs with respect to the galaxy body (i.e., R$_{e, GC}$/R$_e\sim$1) has been found in other UDGs that present no signatures of tidal distortions \citep[see][]{Saifollahi2021}. There is not preferred direction, or axis, in the spatial distribution of GCs of this galaxy. 

However, from Fig. \ref{fig:gc_distribution}, there is a hint that a larger number of GCs are located towards the northern side of the galaxy. To explore whether this excess is statistically significant, we computed its significance by deriving the probability the globular clusters distribute in that way using a binomial distribution. We divided the GCs North and South from the minor axis of the galaxy (PA = $-31$ deg, defined in Sec. \ref{sec:profiles}). 9 (8) clusters are North of the minor axis and 5 (4) are South, for the total (spectroscopically confirmed) sample of GCs. We find a probability of $0.64\pm0.13$ for the total sample and $0.66\pm0.14$ for the spectroscopically confirmed \citep{Shen2020} sample. Within the errors ($\sim1\sigma$), the spatial distribution of the GCs is consistent with the null hypothesis (i.e., the GCs are distributed with similar probability in the two sides of the galaxy, probability of 0.5). Therefore, we can not reject the idea that the observed distribution of GCs in this galaxy is compatible with a random distribution.

\section{The stellar component of \dft}

In this section, we use ultra-deep INT imaging to address whether this galaxy presents tidal features that can point to tidal interactions with one of the nearest massive galaxies and explain its seemingly low content of DM.

\subsection{Surface brightness profiles}\label{sec:profiles}

 \begin{figure*}
 \centering
   \includegraphics[width = 0.9\textwidth]{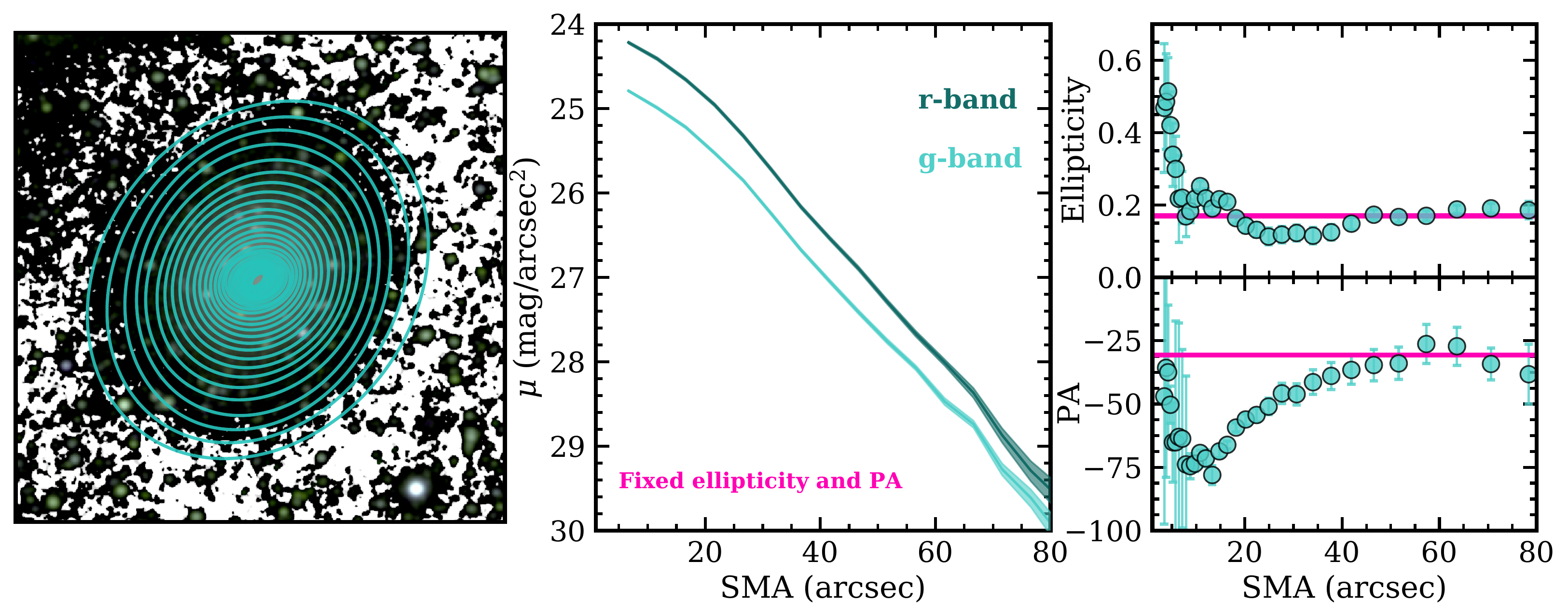}
   \caption{Output from \texttt{ellipse} for the INT $g$ and $r$ images of \dft. The left panel shows the $200\arcsec\times200\arcsec$ region around \dft{} with the fitted ellipses overplotted. The middle panel shows the surface brightness radial profiles as a function of the semi-major axis for the $g$ (mint green) and $r$ (teal). The rightmost panel presents the ellipticity (top) and PA (bottom) radial profiles. The profiles in the middle panel have been derived by fixing the ellipticity and PA to the values indicated as a pink solid line in the rightmost panels.
   \label{fig:profiles}}
    \end{figure*}

The goal of this paper is to investigate whether \dft{} shows any signs of interaction with a massive nearby neighbor. Note that this galaxy's distance has remained controversial and therefore, which massive galaxy it is associated with is unclear. \citet{vD_df2} and \citet{vD2018b} claimed a distance of $\sim20$ Mpc which connects it with NGC 1052\footnote{This same group has claimed a distance of $22.1\pm1.2$ Mpc using deeper \emph{HST} data \citep{Shen2021}.}. On the contrary, \citet{Trujillo2019} and \citet{Monelli2019} place the galaxy at $\sim13.5$ Mpc, associating it with the spiral galaxy NGC 1042\footnote{The revised surface brightness fluctuation estimation in \citet{Zonoozi2021} places an upper limit to the distance to \dft{} of 17 Mpc.}. 

In surface brightness profiles, signs of interactions will appear as an excess of light at large radius and deviations from the morphology of the inner parts of the galaxy \citep[e.g.,][]{Johnston2002}. To explore whether this is the case for \dft, we derived the radial profiles for the INT images using \ellipse. It provides the median intensity, ellipticity and position angle (PA) for each of the fitted isophotes.

Beforehand, we need to mask thoroughly the image. For this, we used the same procedure described in Sec. \ref{sec:rem_star}. In this case, the ``cold” mask was further expanded 6 pixels while the ``hot” was expanded 2 pixels, leaving \dft{} unmasked in the ``cold” mask. That leaves the diffuse light of the galaxy unmasked while all compact sources like background galaxies and GCs are masked. The mask is shown in Appendix \ref{app:mask}.

Once we have all contaminant sources masked, we derived the radial profiles for the INT $g$ and $r$ images with \texttt{ellipse}. In this case, we run \ellipse{} in three steps, similar to the approach followed in \citet{Huang2018}. We first ran the task allowing all parameters to vary freely. In the second run, we fixed the centers to the median centers of the isophotes returned by \ellipse{} in the first iteration. In the third iteration, we hold the ellipticity and PA fixed to derive the surface brightness profiles in both $g$ and $r$ bands. This last step ensures that we reach regions where the signal-to-noise ratio is low.

Fig. \ref{fig:profiles} shows the output of \texttt{ellipse} for the $g$ and $r$ bands. In the leftmost panel, a postage stamp of a $200\arcsec\times200\arcsec$ region around \dft{} is shown with the fitted ellipses overplotted. The right panel shows the ellipticity (top) and PA (bottom) as a function of the semi-major axis (SMA).

The 1-D radial surface brightness profiles as a function of the SMA are shown in the middle panel, to a radius of $\sim80\arcsec$. The shaded regions represent the error in each elliptical isophote. The error is computed as the combination of the error in the intensity of each isophote, as provided by \ellipse, and the error of the sky given by the distribution of background pixels in each image. The 1-D surface brightness profiles in both bands are derived with a fixed ellipticity and PA ($0.17$ and $-31$ deg, respectively, shown as the pink lines in the right panel).

\begin{figure*}
\centering
  \includegraphics[width = 0.4\textwidth]{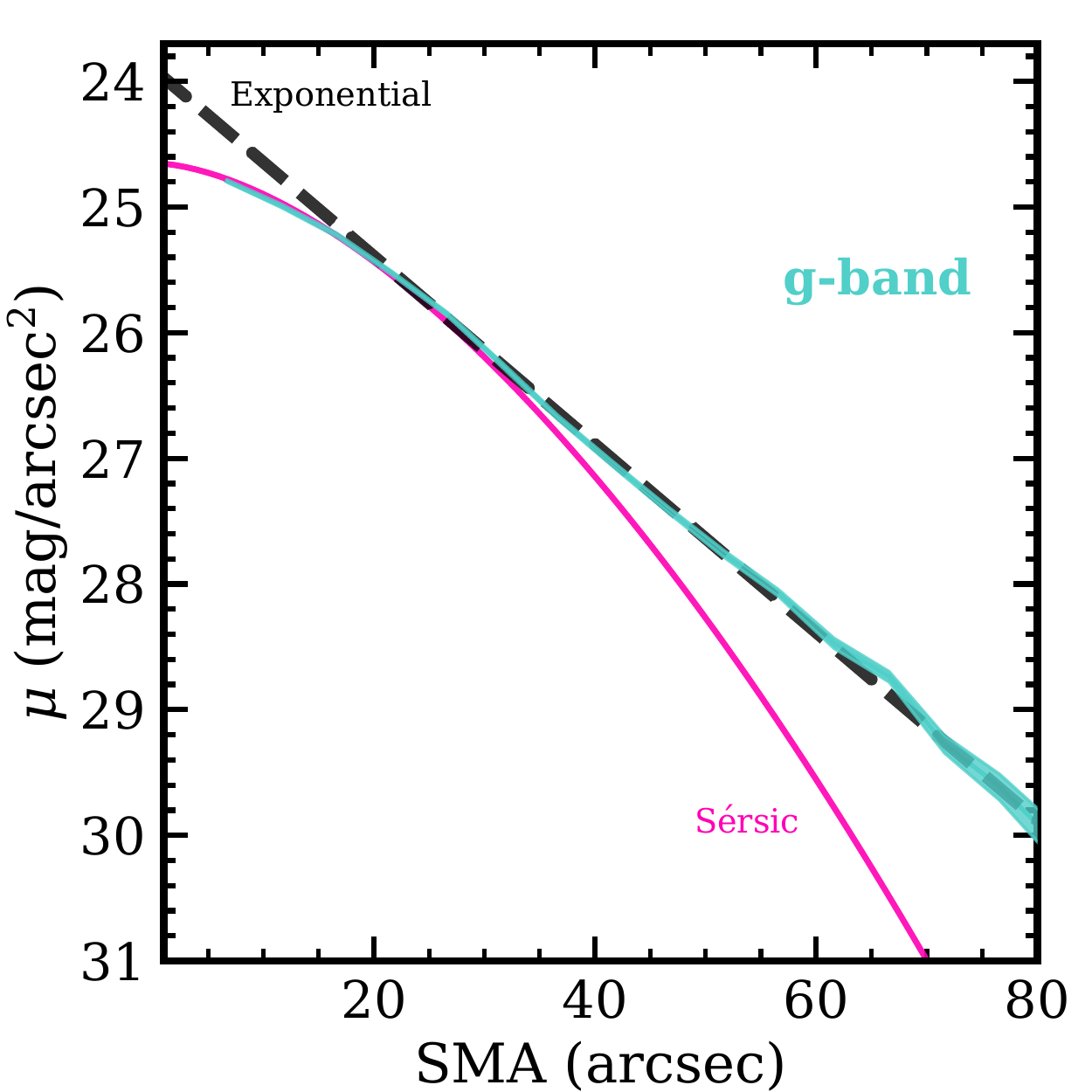}
  \caption{Observed 1-D surface brightness profiles of \dft{} in the g-band (mint green line). The shaded region represents the error in each elliptical isophote computed as described in Sec. \ref{sec:profiles}. The pink solid line shows the S\'ersic fit with n=0.6 and $R_{e} = 22.6$ arcsec as reported in \citet{vD_df2}. The dashed black line correspond to an exponential fit to the profile for $r>20$\arcsec. It is evident that neither the S\'ersic nor an exponential can fully describe the profile of this galaxy. \label{fig:fits}}
    \end{figure*}

These profiles derived for \dft{} show that the structure of the galaxy is more complex than previously reported. Fig. \ref{fig:fits} shows that the observed profile in the $g$-band is well described with a S\'ersic model in the inner $\sim30$ arcsec with n = 0.6. However, this cannot provide a faithful description of the galaxy at larger radius. On the other hand, an exponential fit ($R_{e} = 24.4\pm0.1$ arcsec) cannot reproduce the galaxy in its inner regions. Interestingly, the fact that the outer exponential can not describe the inner regions suggests that if the outer part of the galaxy is a disk structure it is truncated in the central region of the galaxy. Exponentially truncated disks in their inner regions have been proposed by \citet[][]{Breda2020} to explain the structure of many spiral galaxies.

The exponential decline seen in Fig. \ref{fig:profiles} is reinforced by the behaviour of the ellipticity and PA with radius. Once they reach $r\sim35$ arcsec, they become roughly constant. Signs of interaction or a warp in the galaxy will show up as an increase in PA and ellipticity with radius, that are not seen here. The change in the PA with radius seen within $35$ arcsec was also reported in \citet{Muller2019} for this galaxy. From the profiles in Fig. \ref{fig:profiles}, we computed the new effective radius of the galaxy to be $R_e = 28.4\pm0.3$ arcsec.

\begin{figure*}
 \centering
   \includegraphics[width = \textwidth]{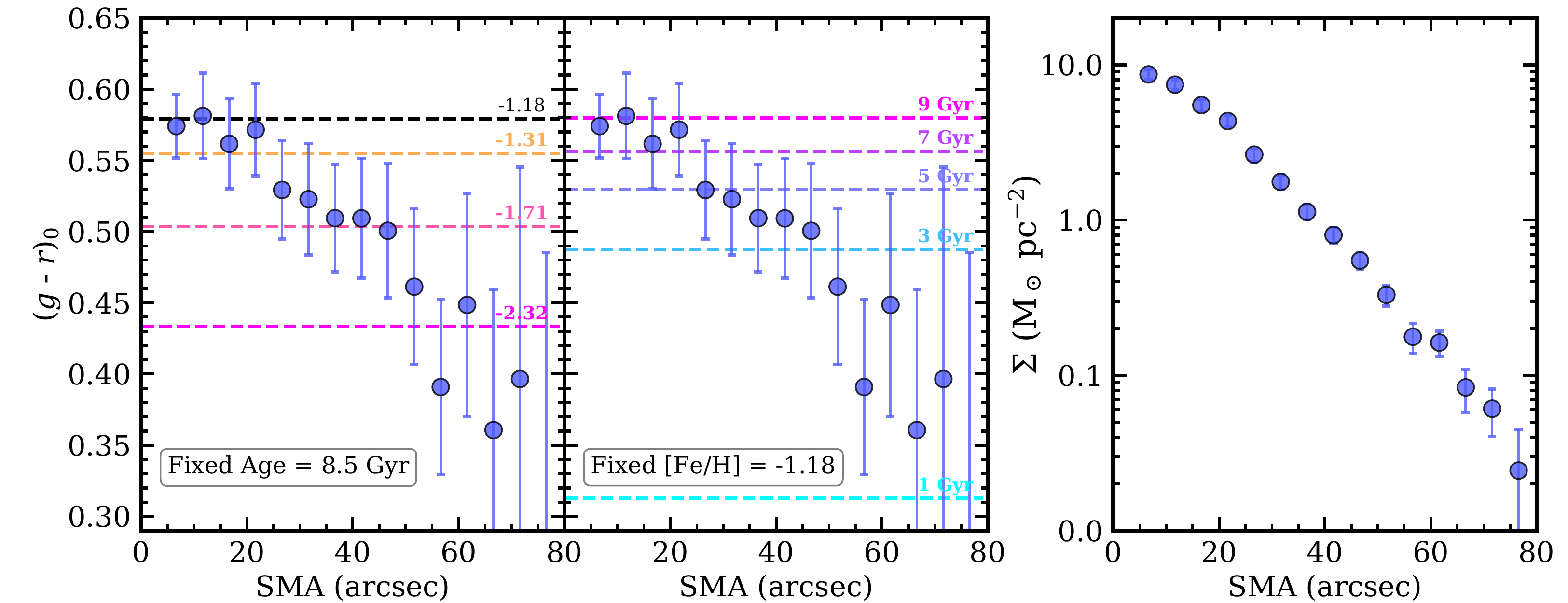}
   \caption{Left and middle panel: the INT $g-r$ color profile of \dft{} shows a clear negative gradient. The horizontal dashed lines in the left panel indicate the $g-r$ color of the \citet{Vazdekis2016} models at a fixed age of $8.5$ Gyr and different [Fe/H] (as labelled), while in the middle panel they indicate different ages at a fixed metallicity of [Fe/H] = -1.18. Right panel: Stellar mass density profile of \dft. \label{fig:color}}
    \end{figure*}

In addition, we derived the radial $g-r$ profile and the stellar mass density profile of \dft. This is shown in Fig. \ref{fig:color}. In the left and middle panels, we plot the $g-r$ color profile to a radius of $70$ arcsec (beyond this radius the errors are larger than 0.2 mag and therefore not reliable). The leftmost panel shows the color estimate for different metallicities derived from the \citet{Vazdekis2016} models for a fixed age of $8.5$ Gyr \citep[i.e., the age of \dft{} estimated spectroscopically,][]{Ruiz2019} as horizontal dashed lines. The black horizontal dashed line corresponds to the metallicity derived for the central part of this galaxy in \citet{Ruiz2019} using spectroscopy.Similarly, the dashed lines in the middle panel are the different colors of the models for a fixed metallicity, [Fe/H] = -1.18, but varying age.
The rightmost panel of Fig. \ref{fig:color} shows the stellar mass density profile for this galaxy. To derive it, we apply equation 1 and 2 in \citet{MT14} (see also \citealt{Bakos2008}) to link the observed surface brightness in the $g$-band to a stellar mass to light (M/L) ratio. The M/L ratio was derived from the prescriptions given in \citet{Roediger2015}. For this, we assumed a \citet{Chabrier2003} initial mass function.

The color profile shows a clear negative gradient with radius. Two different scenarios could explain why \dft{} becomes bluer with radius. In Fig. \ref{fig:color}, we interpret the color gradient as an age and/or a metallicity gradient. We can not rule out any of these two alternatives as it is not possible to disentangle between age and metallicity with only one color. However, the lack of HI detection in this galaxy \citep{Chowdhury2019} suggests that it is similarly old everywhere and that the color trend could be explained as a metallicity gradient along its radial profile. In fact, there is a hint at a metallicity gradient, but no age gradient, in the MUSE data of \dft{} in \citet{Fensch2019}. Assuming that the age of the galaxy is the same along its entire structure, the color gradient would indicate that the stellar population is getting more metal-poor towards the outskirts. If that was the case, the derived metallicity using spectroscopy would only be representative of the innermost parts of the galaxy ($<1R_e$). 

The stellar mass density profile also exhibits the two components, S\'ersic and exponential, present in the surface brightness profiles. There is a shallow core down to $\sim23$ arcsec followed by an exponential decline down to the limit where the data allows to trace reliably the galaxy (i.e. down to $0.07$ M$_{\odot}/\mathrm{pc}^2$).

 \begin{figure*}
 \centering
   \includegraphics[width = 0.9\textwidth]{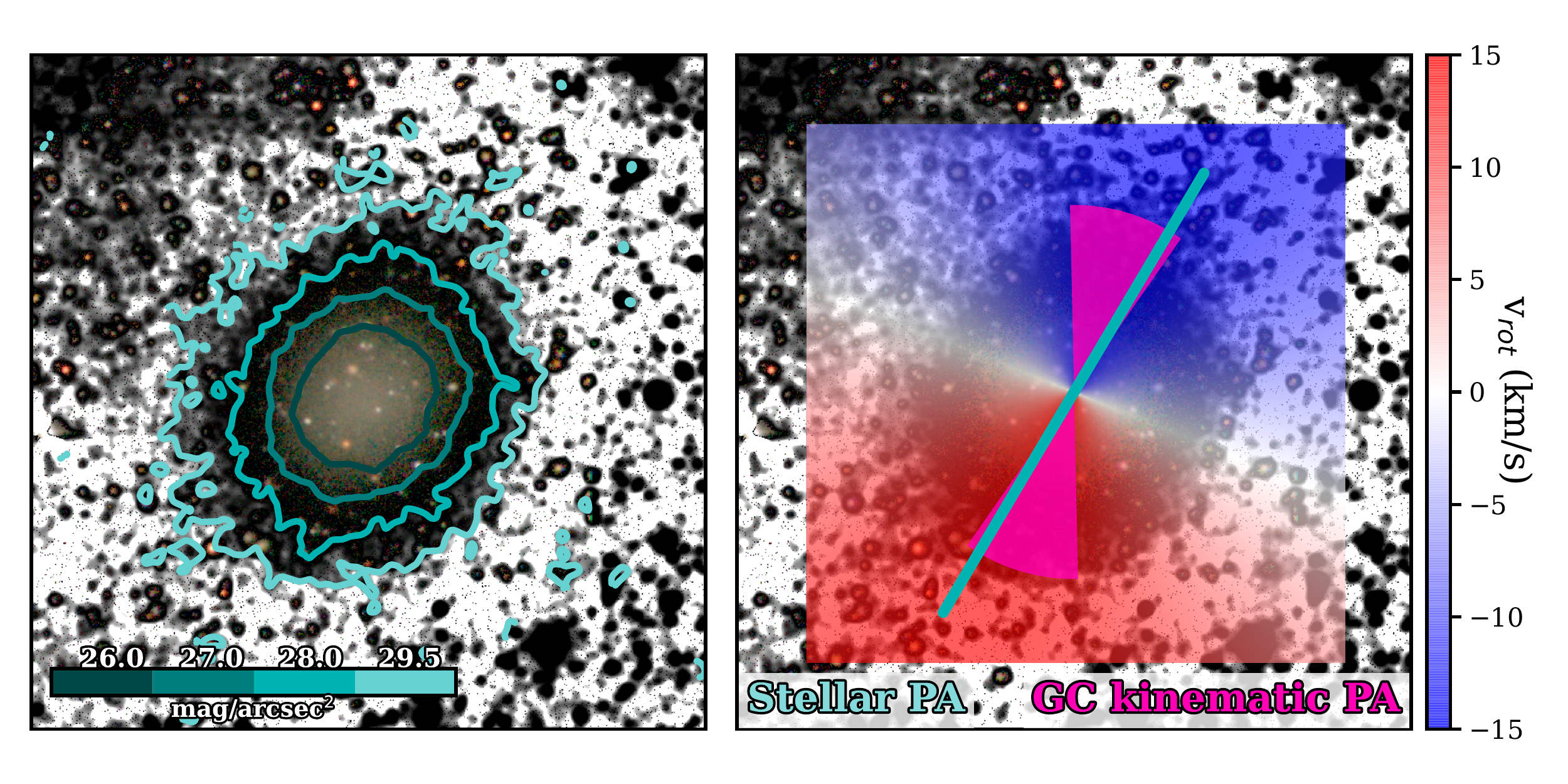}
   \caption{\hipercam{} RGB color image of the $240\arcsec\times240\arcsec$ around \dft{} with the INT $g$ image in black and white for the background. The left panel shows the contours at four different surface brightness levels. There is no evidence of tidal features in our ultra-deep imaging of \dft. The right panel shows the rotation map from \citet{Lewis2020} with its PA (pink area) and the stellar PA from the outer parts. Both PA are in agreement, suggesting that the galaxy is indeed rotating.
   \label{fig:contours}}
    \end{figure*}

\subsection{The shape of \dft}\label{sec:shape}

Matter stripped from satellite galaxies forms tidal tails which will eventually become completely unbound and form larger tidal features. In N-body simulations \citep[e.g., ][]{Klimentowski2009}, dwarf galaxies orbiting a massive host galaxy potential show two tidal tails emanating from the two opposite sides of the dwarf as the particles are seen to move in the direction of the tidal forces, creating a shape reminiscent of an S. 

To explore whether we can rule out the presence of tidal features up to the depth of our data, we derive the isocontours of this galaxy at four different surface brightness limits. For this we use our deepest image: the INT $g$-band. As in \citetalias{Montes2020}, we used \texttt{matplotlib}’s \texttt{contour} to draw the contours in the left panel of Fig. \ref{fig:contours}. We use the same mask derived in Sec. \ref{sec:profiles} and convolve the image with a Gaussian of $\sigma$ = 5 pix (1 pix = $0\farcs33$) to further enhance structures that might be buried in the noise of the outer parts of the galaxy. The different surface brightness levels we explore are: $26.0$ (darker teal shade), $27.0$, $28.0$ and $29.5$ mag/arcsec$^2$ (lightest teal shade).

From the contours in Fig. \ref{fig:contours} we can confirm the results in Sec. \ref{sec:profiles}: the outer parts of \dft{} appear symmetric not showing any sign of tidal stripping down to a surface brightness of 29.5 mag/arcsec$^2$. This image supports what we found for the surface brightness profiles in the previous section: a structure compatible with a low inclination disk.

The limiting surface brightness of the image in the $g$-band (30.4 mag/arcsec$^2$; 3$\sigma$ in $10\arcsec\times10\arcsec$ boxes) can be used to make a rough estimation (upper limit) of the amount of stellar mass that could be hidden under our detection limit. The outer parts of the galaxies have a color $g-r\sim 0.4$. At 30 mag/arcsec$^2$, this is equivalent to 0.03 M$_\odot$/pc$^2$. Therefore, if there were any stellar mass which such density in the radial range between 80 to 100 arcsec (i.e. $\sim$9000 arcsec$^2$), that will correspond to a total stellar mass in that region of $\sim1.1\times10^6$ M$_\odot$ (at 13 Mpc) or $\sim2.5\times10^6$ M$_\odot$ (at 20 Mpc). That is, the amount of potentially stripped stellar mass of DF2 is $\sim2$\% or less of its total stellar mass.

\subsection{Tidal radius of \dft}\label{sec:tidal}

Tidal forces are dependent on the gradient of a gravitational field and so tidal effects are usually limited to the immediate surroundings of a galaxy. The two most prominent galaxies in the FOV, and therefore the ones that could produce tidal stripping in \dft, are NGC 1052 and NGC 1042. \citet{vD_df2} used the vicinity of \dft{} to NGC 1052 ($13\farcm7$ in projection) to support its distance of 20 Mpc. However, \citet{Monelli2019} finds two groups in the same line of sight, one at 13.5 Mpc, containing NGC 1042, and one at $\sim20$ Mpc, containing NGC 1052. They associated \dft{} to the former (NGC 1042), at a projected separation of $20\farcm5$. 

Even though the projected physical distance of \dft{} to both galaxies would be equivalent to $80$ kpc, the total mass of NGC 1052 \citep[M$_{1052,dyn}(<40$ kpc) = $4.14\pm0.71\times 10^{11}$M$_{\odot}$,][]{Forbes2019} is more than 10 times larger than that of NGC 1042 (M$_{1042,dyn}$($<8$ kpc)$= 3.0^{+3.0} _{-1.5}\times10^{10}$ M$_{\odot}$, Trujillo et al. in prep.). In order to estimate the dynamical mass of NGC 1042 to the same radial distance than NGC 1052 (i.e., 40 kpc), we can assume that the rotational velocity of NGC 1042 is constant. Using this, then the dynamical mass is proportional to the radius.  Therefore, the dynamical mass of NGC 1042 (very likely an upper limit due to our assumption) is M$_{1042,dyn}(<40$ kpc$) = 1.5^{+0.3} _{-0.2}\times10^{11}$ M$_{\odot}$. Given this mass difference between the two galaxies, we would expect to see a different tidal effect on \dft{} depending on its associated host.

To explore where we expect to see the tidal features for both NGC 1052 and NGC 1042 we calculated the tidal radius of \dft{} produced by each galaxy. The tidal radius identifies the radius where the tidal effects are expected to be important in the satellite and where we should be looking for tidal distortions. To derive the tidal radius ($r_{t}$) for \dft{}, we follow equation 5 in \citet{Johnston2002} but for the instantaneous $r_{tidal}$; that is the tidal radius at the current position of the satellite from the parent galaxy assuming a circular orbit. This is given by:

\begin{equation}
    r_{tidal,inst} = D \times \left( \frac{m_{dyn}}{M_{host,dyn}} \right)^{1/3}
\end{equation}

\begin{figure*}
\centering
  \includegraphics[width = 0.7\textwidth]{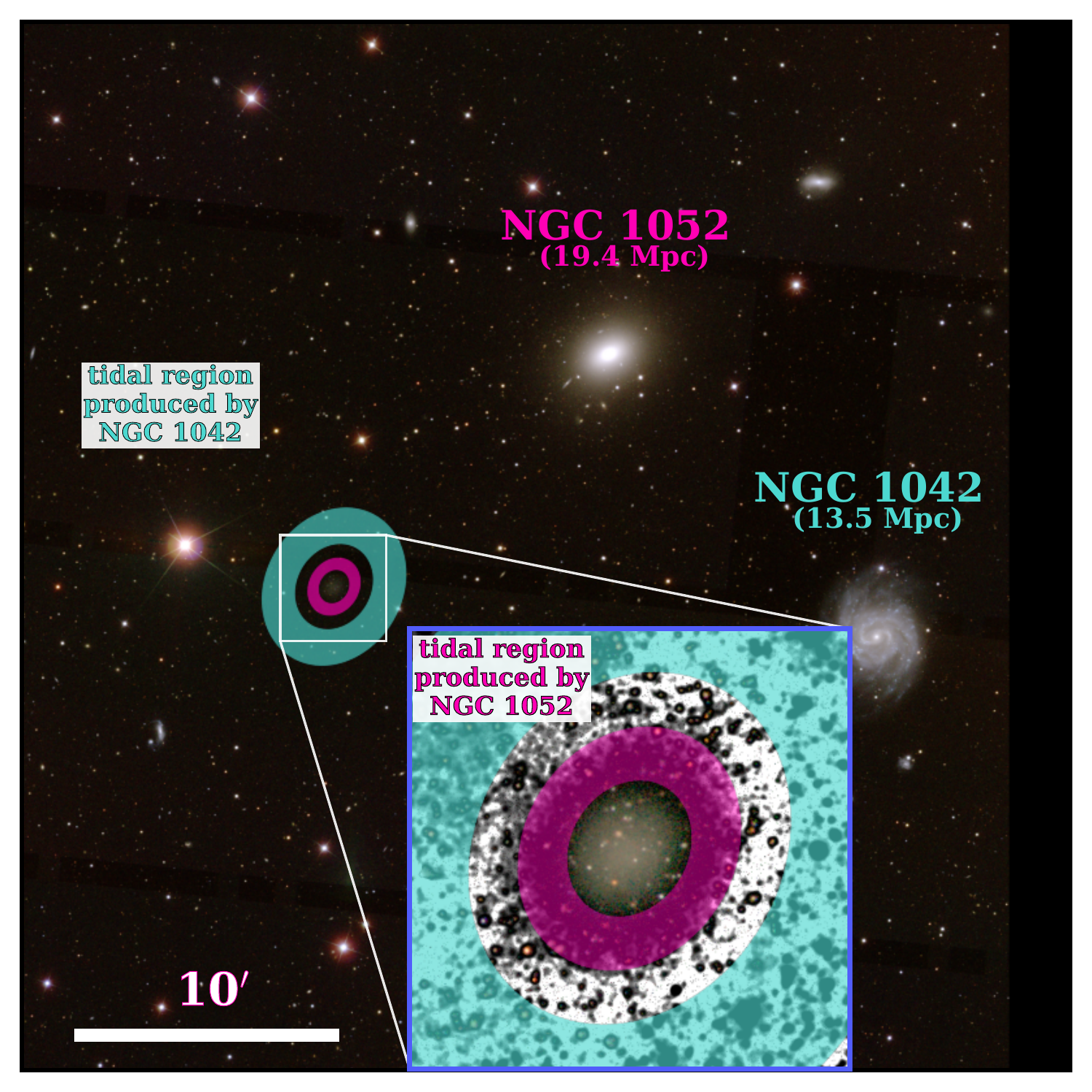}
  \caption{Expected tidal regions on \dft{} produced by NGC 1042 (mint green elliptical ring, bigger) and NGC 1052 (pink elliptical ring, smaller). The background is a SDSS image of the FOV around \dft. The nearby massive galaxies, the elliptical NGC 1052 and the spiral NGC 1042, are labelled. For ease of view, we show a zoom-in of the $240\arcsec\times240\arcsec$ region around \dft. The lack of tidal features around \dft{} does not favour the hypothesis that this UDG is physically bound to NGC 1052.  \label{fig:tidals}}
    \end{figure*}

Fig. \ref{fig:tidals} shows the expected tidal radii derived for NGC 1052 (in pink) and for NGC 1042 (in mint green) as elliptical rings. The elliptical rings encompass the range of radius from the uncertainties in the total masses of the galaxies. It is evident from the Figure that the difference in mass between the two potential hosts will have a different impact in the satellite. The tidal radius produced by NGC 1042 is $r_{t}^{1042}$ = $148_{-48} ^{+39}$ arcsec. On the other hand, the tidal radius if the UDG was associated to NGC 1052 is $r_{t}^{1052}$ = $55_{-17} ^{+14}$ arcsec, well within the observed extension of the galaxy in this work.

\section{Discussion}

In this paper, we study in detail the first galaxy ``lacking" DM, \dft{}, using ultra-deep images to assess if tidal stripping could explain its claimed low content of DM. In \citetalias{Montes2020}, we explored tidal stripping as an alternative scenario to explain the low DM content in \dff{}. Our aim in this paper was to repeat the same analysis for \dft. However, the distribution of GCs (Fig. \ref{fig:gc_distribution}), the surface brightness profiles (Fig. \ref{fig:profiles}) and the contours (left panel of Fig. \ref{fig:contours}) of this galaxy show no signs of interaction. The lack of tidal features implies that currently there is no significant tidal stripping in this galaxy.

\subsection{Evidence of rotation in \dft} \label{sec:rotation}

The dynamical analyses of \dft{} have assumed that this galaxy is pressure supported \citep[e.g., ][]{vD_df2, Martin2018, Laporte2019}. However, if a significant rotation component is present, it will influence the current dynamical mass estimate of this galaxy. 

Dynamical studies of the stellar light of the innermost regions of this galaxy have found contradictory results. The lack of rotation reported in \citet{Danieli2019} is likely caused by the small FOV of the KCWI instrument ($20\farcs4\times16\farcs5$) that coincides with the presence of the bulge-like structure seen in the radial profiles of Fig. \ref{fig:profiles} at $R<35$ arcsec.
In the other hand, evidence for rotation of \dft{} has been presented in \citet{Emsellem2019} and \citet{Lewis2020}. Note that the rotation axis found in the stellar component by \citet{Emsellem2019} is almost perpendicular to the axis found in \citet{Lewis2020} using the GC kinematics. While it is not clear the reason of this disagreement, we speculate that it might be caused by the limited FOV of MUSE ($60\arcsec\times60\arcsec$) imaging only one side of the galaxy compared with the larger radial distances probed with the GCs\footnote{We suggest that the MUSE results should be revisited taking into account the PA of the external parts of the galaxy found in this work.}.

The profiles in Fig. \ref{fig:profiles} show that at $R>35$ arcsec the galaxy presents a component that is best described with an exponential. This suggests the presence of a disk in \dft{} that only shows up in deep images \citep[see also][]{Muller2019}. In the right panel of Fig. \ref{fig:contours}, we show the inner $240\arcsec\times240\arcsec$ region of \dft{} with the rotation map from \citet{Lewis2020} overplotted. The rotation map obtained from re-analyzing the dynamics of the GC system is in nice agreement with the disk structure at large radius. We indicated with the mint line the PA of the outer parts ($R>35$ arcsec) of the galaxy derived from the \ellipse{} fits in Sec. \ref{sec:profiles}. This stellar PA seems to be perpendicular to the projected rotation axis of the rotation map. In order to assess that, we have also plotted the PA of the rotation map and its errors as reported in \citet{Lewis2020} (pink area). Both angles agree within the errors. 

The presence of a disk-like structure in the outer parts of the galaxy provides further evidence of the rotation of \dft. We used eq. 9 in \citet{Lewis2020} to calculate a rough estimate of the dynamical mass of the galaxy considering the rotation component measured in that work. Using the axis ratio of the disk measured with \ellipse{} in Sec. \ref{sec:profiles}, we can derive a rough estimate of the inclination angle $i$ of this galaxy using $b/a = cos(i)$ (here we do not consider any thickness for the disk). That gives us an inclination angle of $34^{+6}_{-7}$ degrees. The rotation parameters were taken from the analysis in \citet{Lewis2020}: $v_{rot} = 12.44^{+4.40}_{-5.16}$ km/s and $\sigma = 3.88^{+4.42}_{-2.72}$ km/s. 

Under the assumption that the galaxy is at 20 Mpc, we found that the dynamical mass enclosed by the GCs is M($<7.6$ kpc) = $9.0^{+17.1} _{-6.8}\times 10^8$ M$_{\odot}$. Assuming a stellar mass of $\sim 2\times 10^8$ M$_{\odot}$, this corresponds to M$_{dyn}$/M$_*$ $= 4.5^{+8.6} _{-3.4}$. If a distance of 13.5 Mpc is assumed \citep{Trujillo2019, Zonoozi2021}, then the estimates are : M($<5.1$ kpc) = $6.1^{+11.5} _{-4.5}\times 10^8$ M$_{\odot}$ and  M$_{dyn}$/M$_*$ $ = 12.1^{+23.0} _{-9.1}$, assuming a stellar mass of $5\times10^7$ M$_{\odot}$ \citep{Trujillo2019}. The dynamical mass derived here is a factor of 2.6 larger than that derived in \citet{vD_df2} for a distance of 20 Mpc, and a factor of 1.5 larger than in \citet{Trujillo2019} at 13.5 Mpc.

As discussed in \citet{Lewis2020}, the presence of rotation in this galaxy evidenced also by the presence of a disk component at large radius, affects the inferred M/L ratio bringing the DM content of \dft{} more in line with galaxies of similar stellar masses \citep[e.g.,][]{Mancera2019, Mueller2020}. Furthermore, a closer distance of $\sim13.5$ Mpc would make even more room for DM in this galaxy, and far from ``lacking" dark matter.

\subsection{The tidal radius of \dft{} and its distance}\label{sec:disc_tidal}

In Sec. \ref{sec:tidal}, we discuss the estimated tidal radii of \dft{} depending on the two potential host galaxies: NGC 1042 at 13.5 Mpc \citep{Trujillo2019} or NGC 1052 at 20 Mpc \citep{vD_df2}. Both radii were visually represented in Fig. \ref{fig:tidals}.

Our results suggest that if \dft{} was gravitationally bound to NGC 1052, we would have already seen tidal features around this UDG. However, the lack of any sign of tidal stripping is not in contradiction with a potential association of \dft{} with NGC 1042 and, therefore, it is consistent with the distance of $13.5$ Mpc found in \citet{Trujillo2019} and \citet{Monelli2019}. One can argue that we are observing the projected distance between \dft{} and its host, which would be the minimum distance between both galaxies. The fact that we are not detecting any feature within a radius of $80$ arcsec means that this would be the minimum tidal radius. That is a factor of $\sim2$ larger than the $r_{t}^{1052}$ calculated in Sec. \ref{sec:tidal} and therefore we can infer that the distance between both galaxies has to be larger than $\sim168$ kpc. The virial radius of NGC 1052 is $390$ kpc \citep{Forbes2019}, and consequently, we cannot fully reject the scenario where \dft{} and NGC 1052 are associated only on the basis of the absence of tidal features. However, the lack of tidal distortion is indicative that if they are associated, \dft{} is in its initial infall into the group. If this was not the case, \dft{} would have already experienced the tidal forces of the massive galaxy or even been disrupted. This is in agreement with \citet{Forbes2019}. The idea that the absence of tidal features indicates an initial infall of the UDG also holds if we use the dynamical mass derived in the previous section (Sec. \ref{sec:rotation}). A higher dynamical mass (and therefore a larger amount of dark matter) will ``shield" the UDG from the tidal forces produced by NGC 1052. The new tidal radius using the new dynamical mass estimate would be $(2.6)^{1/3}\times r_{t}^{1052} = 75^{+19}_{-24}$ arcsec, at the limit of our ultra-deep imaging.

\section{Conclusions}

The existence of long-lived ($\sim$9 Gyr) and stable (without signatures of tidal distortion) galaxies ``lacking" DM present a challenge in our understanding of galaxy formation. This is especially problematic if the galaxy inhabits a dense environment. Here, we explore the tidal stripping scenario to explain the claimed low content of DM in the first of this kind of galaxies: \dft. We use ultra-deep data to image the outer parts of the galaxy down to $R\sim 80$ arcsec (i.e., three times larger than previous imaging of this galaxy) finding no signs of tidal features. However, we see a disk-like structure from $35$ to $80$ arcsec in agreement with the rotation previously reported for this galaxy. The dynamical mass derived taking into account the rotation component is a factor 2.6 larger than in \citet{vD_df2} and 1.5 larger than the one measured in \citet{Trujillo2019}. This translates into a dark matter vs. stellar mass ratio of $4.5^{+8.6} _{-3.4}$ at $20$ Mpc ($R<7.6$ kpc) or $12.1^{+23.0}_{-9.1}$ at $13.5$ Mpc ($R<5.1$ kpc), bringing the DM content of this galaxy more in line with other galaxies of similar stellar masses \citep[ $\gtrsim 10$,][]{Mancera2019, Mueller2020}.


\acknowledgments
We thank the referee for their useful comments that helped improve the original manuscript. We would like to thank Javier Rom\'an for his useful comments on the paper and his reduction of the INT images. We also thank Geraint Lewis for providing the rotation map of \dft{} and helpful comments about the rotation of this system. 
We acknowledge support from the grant PID2019-107427GB-C32 from the Spanish Ministry of Science and Innovation. We acknowledge financial support from the European Union's Horizon 2020 research and innovation programme under Marie Sk\l odowska-Curie grant agreement No. 721463 to the SUNDIAL ITN network, and the European Regional Development Fund (FEDER), from IAC project P/300624, financed by the Spanish Ministry of Science, Innovation and Universities (MCIU), through the State Budget and by the Canary Islands Department of Economy, Knowledge and Employment, through the Regional Budget of the Autonomous Community.

A. B. was supported by an appointment to the NASA Postdoctoral Program at the NASA Ames Research Center, administered by Universities Space Research Association under contract with NASA. Based on observations made with the GTC telescope, in the Spanish Observatorio del Roque de los Muchachos of the Instituto de Astrof\'isica de Canarias, under Director’s Discretionary Time. Based on observations made with the Isaac Newton Telescope, in the Spanish Observatorio del Roque de los Muchachos of the Instituto de Astrof\'isica de Canarias.

\facilities{HST(ACS), INT, GTC}


\software{Astropy \citep{Astropy2018},  
          \sextractor{} \citep{Bertin1996},
          \scamp{} \citep{Bertin2006},
          \swarp{} \citep{Bertin2010},
          Gnuastro \citep{Akhlaghi2015},
          Maneage \citep{Akhlaghi2020},
          \texttt{photutils} v0.7.2 \citep{Bradley2019},
          \texttt{pillow} \citep{pillow2020},
          \texttt{ellipse} \citep{Jedrzejewski1987},
          \texttt{numpy} \citep{oliphant2006},
          \texttt{scipy} \citep{scipy2020},
          \texttt{Astrodrizzle} \citep{Gonzaga2012},
          \texttt{Astrometry.net} \citep{Lang2010}}

\vspace{5cm}
\null\newpage

\appendix 
\section{Masking of \dft}\label{app:mask}
\begin{figure}
\centering
  \includegraphics[width = 0.4\textwidth]{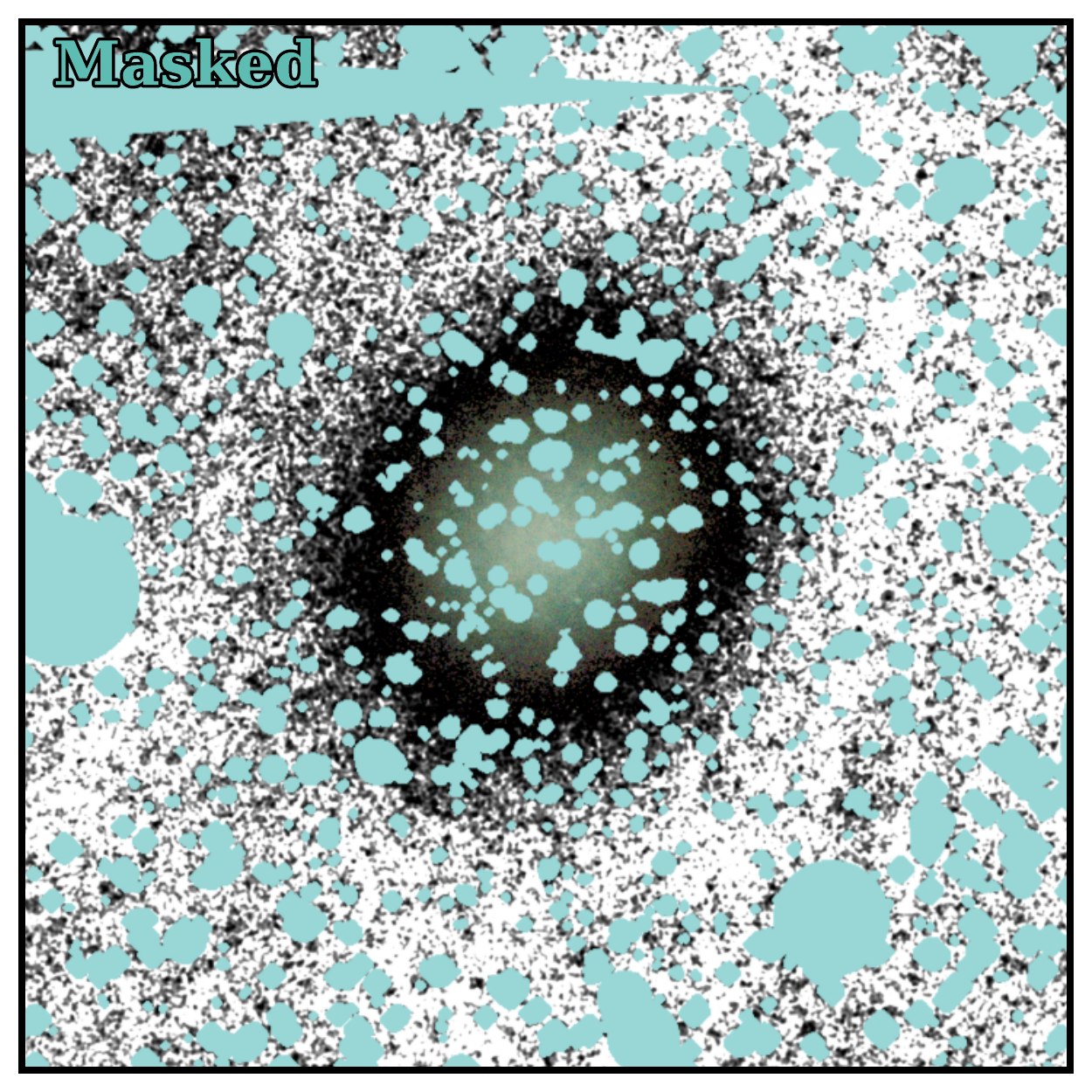}
  \caption{Mask (teal regions) applied to a \hipercam{} color composite of a postage stamp of a region $140\arcsec \times 240 \arcsec$ around \dft. The black and white background is the $g$-band INT image. This shows the thorough masking necessary in these deep images. \label{fig:mask}}
    \end{figure}

Sec. \ref{sec:profiles} describes the procedure to create the mask that is used to derive the surface brightness profiles. Fig. \ref{fig:mask} shows this mask applied to a composite of an RGB \hipercam{} images and a black and white background from the $g$-band INT image.

\bibliography{df2}{}
\bibliographystyle{aasjournal}

\end{document}